\documentclass[%
 preprint, 
 amsmath,amssymb,
 aps, physrev,
]{revtex4-2}

\usepackage{subcaption}

\usepackage{graphicx}
\usepackage{dcolumn}
\usepackage{bm}
\usepackage{braket} 
\usepackage{hyperref}

\begin{document}
\setlength{\parskip}{0pt}


\title{\textbf{Environment-assisted transport in a strongly correlated boundary-driven Fermi-Hubbard chain} 
}%

\author{Hakan Yeler}
\email{Contact author: hakanyeler@iyte.edu.tr}
 \altaffiliation[]{}
\author{Özgür Çakır}%
 \email{Contact author: ozgurcakir@iyte.edu.tr}
\affiliation{%
Physics Department, Izmir Institute of Technology, 35430, Urla, Izmir, Turkey
}%



\date{\today}

\begin{abstract}
We study steady-state transport in a one-dimensional Fermi-Hubbard chain
coupled to particle reservoirs at the boundaries and to local dephasing
baths at each site, using the time-evolving block decimation (TEBD)
method to solve the Lindblad master equation. In the absence of dephasing, the current exhibits two well-separated
maxima as a function of the boundary driving rate, reflecting the
distinct charge and spin energy scales of the strongly correlated
regime. Upon introducing dephasing, we find two
distinct dephasing-induced transport-enhancement regimes, in contrast to
the single enhancement previously reported for spinless fermions.
Analysis of the non-equilibrium steady state in the Hamiltonian eigenbasis
reveals that the two regimes originate from distinct dephasing-induced
redistribution processes: the first involves redistribution within the
uppermost Hubbard band, while the second involves transitions between
Hubbard bands. Our results demonstrate how many-body correlations shape
the interplay between coherent driving, dephasing, and quantum Zeno
physics in strongly correlated open systems.
\end{abstract}

\maketitle


\section{\label{sec:intro}INTRODUCTION}

Understanding transport in open quantum systems is a central problem in nonequilibrium quantum physics, motivated by both fundamental questions and potential applications in nanoscale quantum technologies. In realistic quantum devices, the unavoidable coupling between a system and its environment strongly influences transport properties. While environmental noise is generally expected to suppress coherent transport, it has become clear that under suitable conditions it can instead enhance transport efficiency. This phenomenon, known as environment-assisted quantum transport (ENAQT), has been extensively studied in single-particle systems, including disordered lattices, biological light-harvesting complexes, and photonic networks \cite{enaqt_mobilityedge,fleming_fotosentez,enaqt_ordered,León-Montiel_Quiroz-Juárez_Quintero-Torres_Domínguez-Juárez_Moya-Cessa_Torres_Aragón_2015,enaqt_10qubitnetwork,Lloyd_fotosentez,Viciani_Gherardini_Lima_Bellini_Caruso_2016a,enaqt_mechanism,NESSbio,ENAQT}. In these systems, dephasing reduces destructive quantum interference, allowing particles to access more delocalized states and thereby enhancing transport.

The interplay between dissipation and many-body correlations is considerably richer and remains far less understood. Open interacting quantum systems have attracted growing theoretical and experimental attention in recent years, including realizations with ultracold atoms, quantum simulators, and engineered solid-state platforms \cite{review_theoretical,review_experiment,experiment_coldgas,expansionofcoldatom,experiment_transport,ultracold1,experiment_2,exp1}. In boundary-driven one-dimensional strongly correlated systems such as the Hubbard and $t$-$V$ models, strong driving may produce negative differential conductivity (NDC) through the formation of domain-like configurations that inhibit transport \cite{PhysRevB.80.035110}. Building on this picture, bulk dephasing was shown to enhance transport in strongly interacting spinless fermion chains by inducing incoherent transitions from high-energy dark states into more mobile many-body states \cite{PhysRevB.87.235130}.

Whether the same mechanism persists in spinful strongly correlated systems remains largely unexplored. In contrast to spinless models, the interplay between spin degrees of freedom and on-site interactions in the Fermi-Hubbard model generates a substantially richer many-body spectrum and more intricate transport pathways. These additional degrees of freedom naturally raise the question of whether dephasing-assisted transport follows the same mechanism or gives rise to qualitatively new behavior.

In this work, we investigate steady-state transport in a boundary-driven Fermi-Hubbard chain subject to local dephasing. We show that strong interactions give rise to two distinct dephasing-assisted transport enhancement regimes, which we relate to intra-band redistribution and inter-band transitions in the Hamiltonian eigenbasis, respectively, in contrast to the single enhancement previously reported for spinless fermions \cite{PhysRevB.87.235130}. We further show that, even in the absence of dephasing, the boundary driving rate alone induces a non-monotonic current with two well-separated maxima, whose positions reflect the distinct charge and spin energy scales of the strongly correlated regime.

To capture these effects, we describe the open-system dynamics within the Lindblad formalism \cite{Lindblad_1976b,Breuer_Petruccione_2010}, adopting
a phenomenological description with local boundary driving and dephasing. Although microscopic derivations generally lead to nonlocal dissipative terms, local Lindblad operators have been shown to provide an accurate effective description of boundary-driven transport \cite{micro_deriv}.

The remainder of this paper is organized as follows. In Sec.~\ref{sec:model}, we introduce the model and numerical methods. Sec.~\ref{sec:driving} examines the dependence of the steady-state current on the driving rate. Sec.~\ref{sec:dephasing} presents the effect of dephasing and the microscopic mechanism underlying dephasing-assisted transport. Finally, Sec.~\ref{sec:conclusion} summarizes our conclusions.

\section{\label{sec:model}MODEL AND METHODS}

\begin{figure}[tb]
    \centering
    \includegraphics[width=.6\textwidth]{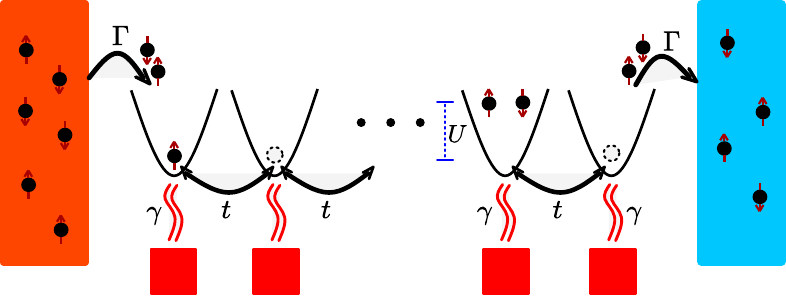}
    \caption{\label{fig:model}Schematic representation of the boundary-driven Fermi-Hubbard chain with local dephasing.}
\end{figure}

We consider a one-dimensional Fermi-Hubbard chain with open boundary conditions, schematically illustrated in Fig.~\ref{fig:model}. The system Hamiltonian is
\begin{equation}
\hat{H}
=
-t\sum_{i=1}^{N-1}\sum_{\sigma}
\left(
\hat{b}_{i,\sigma}^{\dagger}\hat{b}_{i+1,\sigma}
+
\hat{b}_{i+1,\sigma}^{\dagger}\hat{b}_{i,\sigma}
\right)
+
U\sum_{i=1}^{N}
\hat{n}_{i,\uparrow}\hat{n}_{i,\downarrow},
\end{equation}
where $\hat{b}_{i,\sigma}^{\dagger}$ ($\hat{b}_{i,\sigma}$) creates (annihilates) a fermion with spin $\sigma$ on site $i$, $t$ is the nearest-neighbor hopping amplitude, and $U$ is the on-site interaction strength. We consider repulsive interactions ($U\ge0$) and refer to a
site with two fermions ($\hat{n}_{i\uparrow}=\hat{n}_{i\downarrow}=1$)
as a doublon.

The coupling between the system and its environment is assumed to be Markovian, such that the density matrix evolves according to the Lindblad master equation (setting $\hbar=1$),

\begin{equation}\label{eq:Lindblad}
\dot{\rho}
=
\mathcal{L}\rho
=
-i[\hat{H},\rho]
+
\mathcal{D}_{\mathrm{inj}}[\rho]
+
\mathcal{D}_{\mathrm{ext}}[\rho]
+
\mathcal{D}_{\mathrm{deph}}[\rho],
\end{equation}
where the Lindblad dissipator is
\begin{equation}
\mathcal{D}[\rho]
=
\sum_k
\left(
\hat{L}_k\rho\hat{L}_k^\dagger
-
\frac12
\left\{
\hat{L}_k^\dagger\hat{L}_k,\rho
\right\}
\right),
\end{equation}
with $\hat{L}_k$ denoting the quantum-jump operators describing the system-environment coupling.

Particle transport is generated by two reservoirs coupled locally to the
boundary sites, which act as a particle source at site $1$ and a particle
sink at site $N$. Assuming equal injection and extraction rates $\Gamma$,
the corresponding Lindblad operators are
\[
\hat{L}_{\sigma}^{\mathrm{inj}}
=
\sqrt{\Gamma}\,
\hat{b}_{1,\sigma}^{\dagger},
\qquad
\hat{L}_{\sigma}^{\mathrm{ext}}
=
\sqrt{\Gamma}\,
\hat{b}_{N,\sigma}.
\]
This corresponds to the maximal-bias limit, in which particles are only
injected at site $1$ and only extracted at site $N$.

In addition, each lattice site is coupled to an independent local dephasing bath through

\[
\hat{L}_{i}^{\mathrm{deph}}
=
\sqrt{\gamma}\,
\hat{n}_{i},
\]
where
\[
\hat{n}_{i}
=
\hat{n}_{i,\uparrow}
+
\hat{n}_{i,\downarrow}
\]
is the total occupation operator on site $i$, and $\gamma$ is the dephasing rate. This form of dephasing preserves the local particle number while suppressing coherences between different occupation configurations.

The nonequilibrium steady state (NESS) is defined by
\begin{equation}\label{eq:LindbladNESS}
\mathcal{L}\rho_{\mathrm{NESS}}=0.
\end{equation}
Transport is characterized by the steady-state particle current. Using the lattice continuity equation, the current is given by
\begin{equation}
\langle J\rangle
=
it\sum_{\sigma}
\left\langle
\hat{b}_{j+1,\sigma}^{\dagger}\hat{b}_{j,\sigma}
-
\hat{b}_{j,\sigma}^{\dagger}\hat{b}_{j+1,\sigma}
\right\rangle,
\end{equation}
where $1\leq j<N$ and $\sigma=\uparrow,\downarrow$.

Since the Hilbert-space dimension grows exponentially with system size, direct solution of Eq.~\eqref{eq:LindbladNESS} rapidly becomes computationally prohibitive. Instead, the NESS is obtained by evolving the Lindblad dynamics in time using the TEBD algorithm \cite{opentebd} until convergence. The density matrix is vectorized using the superfermion representation \cite{superfermion}, which preserves the fermionic anticommutation relations throughout the time evolution. Results for $N=4$ are obtained by exact diagonalization, while all larger systems are simulated using the ITensor library \cite{itensor}.

The TEBD simulations employ an SVD truncation cutoff of
$\epsilon=10^{-7}$, a maximum bond dimension $\chi=200$, and a Trotter
time step $\tau t=0.1$. The steady state is identified once (i) the
relative change in the current between successive time steps falls below
$10\epsilon$, and (ii) the spatial variation of the current over all
bonds is less than $20\%$ of its mean value, as expected in the NESS. For
$N=4$, the TEBD results agree with exact diagonalization with an average
relative deviation of $0.87\%$ over the range of driving and dephasing
rates considered in this work. The maximum deviation is $9.9\%$, occurring
at the strongest driving rate ($\Gamma/t=100$), where convergence is most
challenging due to the onset of the quantum Zeno regime.

For the most extreme parameter values, where reaching full steady-state
convergence required prohibitively long simulation times, the current was
obtained by rescaling the nearest converged result using the known
asymptotic scaling. In the quantum Zeno regime, the current follows
$\langle J\rangle \propto 1/\Gamma$ ($1/\gamma$), a behavior that was
independently verified by exact diagonalization for $N=4$. For example,
the value at $\Gamma/t=1000$ was obtained from the converged TEBD result
at $\Gamma/t=100$ via
\[
J(1000)=J(100)\times\frac{100}{1000}.
\]
Likewise, in the weak-driving regime, where $\langle J\rangle\propto
\Gamma$, the same rescaling procedure was applied to obtain the current
at $\Gamma/t=10^{-3}$ from the nearest converged point.

For comparison, we also solve the same Lindblad equation, Eq.~\eqref{eq:Lindblad}, within a self-consistent Hartree-Fock mean-field approximation. The interaction term is decoupled as $U\hat{n}_{i\uparrow}\hat{n}_{i\downarrow} \rightarrow U\left(\langle \hat{n}_{i\uparrow}\rangle\hat{n}_{i\downarrow} + \hat{n}_{i\uparrow}\langle\hat{n}_{i\downarrow}\rangle \right)$, rendering the Hamiltonian quadratic, so that the Lindblad dynamics closes at the level of the single-particle correlation matrix $C_{ij\sigma} = \langle\hat{b}_{i\sigma}^{\dagger}\hat{b}_{j\sigma}\rangle$. The steady state is obtained by iterating the correlation-matrix equations until the local densities entering the mean-field Hamiltonian reach self-consistency.

\section{\label{sec:driving}Boundary Drive Rate and Current Resonances}
\begin{figure}[htb]
  \includegraphics[width=.5\textwidth]{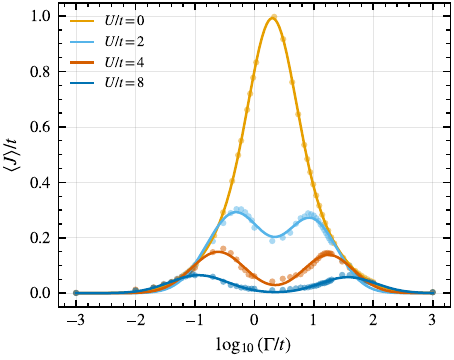}
\caption{\label{fig:jvsrate}Current $\langle J\rangle$ as a function of driving rate $\Gamma$ for an
$8$-site system without dephasing $\gamma/t = 0$, shown for different interaction
strengths $U$.} 
\end{figure}
In our model, the boundary drive rate $\Gamma$ is the sole parameter characterizing the coupling to the particle reservoirs. Fig.~\ref{fig:jvsrate} shows the steady-state current as a function of $\Gamma$ for several interaction strengths $U$.

The first striking feature is the qualitative difference between the non-interacting and interacting regimes. For $U=0$, the current initially increases with $\Gamma$, reaches a maximum at $\Gamma=2t$, and then decreases under stronger driving. While a classical rate-equation picture would predict saturation once the injection rate exceeds the intrinsic transport rate, the observed decrease is a purely quantum effect. Strong boundary driving continuously projects the boundary sites, suppressing coherent transport through the quantum Zeno effect and thereby reducing the current.
\begin{figure}[htbp]
\centering
\includegraphics[width=\textwidth]{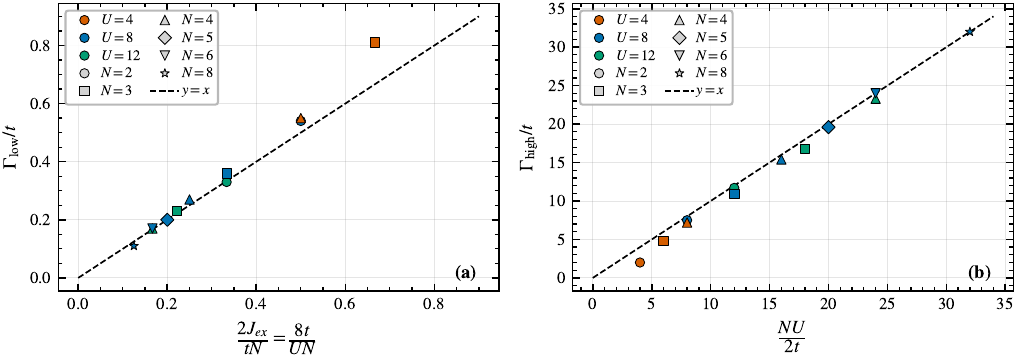}
\caption{\label{fig:driving_scaling}
Scaling of the two current resonances with system size $N$ and
interaction strength $U$, for $N=2$--$8$ and $U/t=4,8,12$. (a)~
$\Gamma_{\mathrm{low}}$ versus $2J_{\mathrm{ex}}/N=8t^2/UN$. (b)~
$\Gamma_{\mathrm{high}}$ versus $NU/2$. Dashed lines indicate $y=x$.}
\end{figure}
For finite interaction strengths, the current exhibits two well-separated resonances. To identify the origin of the two resonances, we next examine their scaling with system size and interaction strength. Numerically, we find that their positions follow
\begin{equation*}
\Gamma_{\mathrm{low}}\sim\frac{2J_{\mathrm{ex}}}{N},\qquad
\Gamma_{\mathrm{high}}\sim\frac{NU}{2},
\end{equation*}
where $J_{\mathrm{ex}}=4t^{2}/U$ is the superexchange energy, as shown in
Fig.~\ref{fig:driving_scaling}. These two characteristic scales differ by a factor of $N^{2}U^{2}/16t^{2}\gg1$ in the strongly correlated regime, reflecting the large separation between the charge and spin energy scales near half filling. 

Since Fig.~\ref{fig:driving_scaling} is obtained at fixed $t$, the two
exponents entering $J_{\mathrm{ex}}$ are not resolved separately there.
Exact diagonalization of short chains separates them: varying the hopping over $t\in[0.7,1.4]$ the low-drive maximum follows
$\Gamma_{\mathrm{low}}\propto t^{2.03}$, whereas the high-drive maximum
is independent of $t$ to within $3\%$. The dimensionless ratios
$\Gamma_{\mathrm{low}}N/J_{\mathrm{ex}}$ and $\Gamma_{\mathrm{high}}/(NU/2)$
are constant to within a few percent over the parameter ranges studied.
The two resonances therefore scale oppositely in all three parameters,
$\Gamma_{\mathrm{low}}\propto t^{2}/UN$ against
$\Gamma_{\mathrm{high}}\propto t^{0}UN$.

By contrast, the mean-field solution predicts size-independent peak positions,
\[
\Gamma_{\mathrm{MF,low}}\sim\frac{2t^{2}}{U},\qquad
\Gamma_{\mathrm{MF,high}}\sim U,
\]
indicating that the observed system-size dependence originates from many-body correlations beyond mean field.

The distinct scalings suggest that the two resonances have different
physical origins. The high-$\Gamma$ resonance sits at the interaction
scale $U$, where real doublon--holon excitations are created, while
the low-$\Gamma$ resonance sits far below it at the superexchange
scale $J_{\mathrm{ex}}=4t^{2}/U$, reflecting the separation between
the charge and spin energy scales near half filling. The physical
origin of the two resonances is discussed below by following the
evolution of the steady state as the boundary driving is increased.

\begin{figure}[htbp]
\centering
\includegraphics[width=\textwidth]{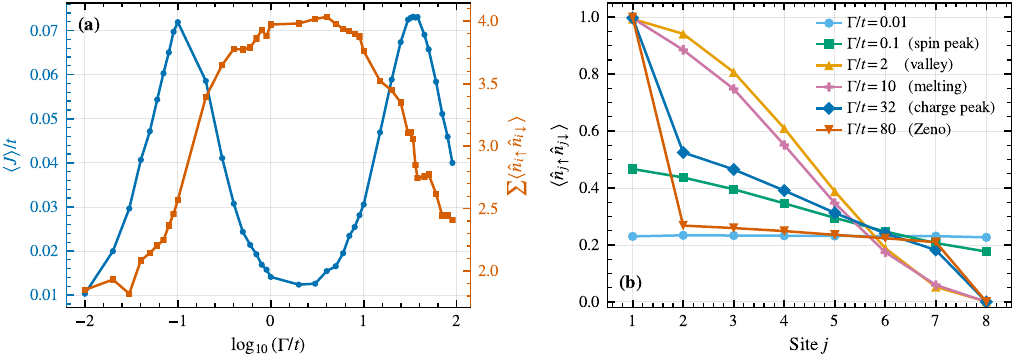}
\caption{\label{fig:dublonprofil}(a)~Steady-state current $\langle J\rangle$ (left axis) and total doublon number $\sum_j \langle n_{j\uparrow}n_{j\downarrow}\rangle$ (right axis) as functions of the boundary driving rate $\Gamma$, for an $N=8$ chain with $U/t=8$. The current exhibits two resonances, separated by a suppressed-current regime in which the doublon number is maximal. (b)~Site-resolved doublon density $\langle n_{j\uparrow}n_{j\downarrow}\rangle$
for representative driving rates spanning the low-drive resonance
($\Gamma/t=0.1$), the suppressed-current valley ($\Gamma/t=2$), an
intermediate melting regime ($\Gamma/t=10$), the high-drive resonance
($\Gamma/t=32$), and the quantum Zeno regime ($\Gamma/t=80$), together
with a weak-driving reference ($\Gamma/t=0.01$).}
\end{figure}

\paragraph*{Low-drive resonance.}
The first resonance occurs at
\begin{equation}
\Gamma_{\mathrm{low}}\sim\frac{2J_{\mathrm{ex}}}{N},
\end{equation}
well below the charge-excitation scale $U$. At weak driving the
current grows linearly, $\langle J\rangle\propto\Gamma$, while the
bulk profile remains nearly flat. As $\Gamma$ increases the profile
becomes linear and the current reaches its maximum near the onset of
the subdiffusive regime, where the profile approaches the cosine form
of high-bias
limit~\cite{Hightempdiffusive} and the current
decreases.

That $J_{\mathrm{ex}}$ is not a generic strong-coupling scale is
confirmed by our own exact diagonalization comparison with spinless fermions subject to nearest-neighbor repulsion $V$:  $t^{2}/V$ is equally available as a candidate scale there, yet we find numerically that the low-drive maximum instead scales as $V^{-2}$. Consistently, no doublon-rich domain
forms at $\Gamma_{\mathrm{low}}$ (Fig.~\ref{fig:dublonprofil}), confirming
that the resonance is not dominated by real doublon--holon excitations.

\paragraph*{Suppressed-current regime.} As the boundary driving is increased beyond the low-drive resonance, charge rearrangement becomes progressively more pronounced and the current enters a strongly suppressed regime. The corresponding steady state exhibits a robust doublon-rich domain occupying approximately the left half of the chain, \begin{equation} \sum_{j}\langle \hat{n}_{j\uparrow}\hat{n}_{j\downarrow}\rangle \simeq \frac{N}{2}. \end{equation} The emergence of this correlated domain is closely related to the mechanism underlying negative differential conductance identified in Ref.~\cite{PhysRevB.80.035110}, where a ferromagnetic domain forms near the driven boundary and blocks transport. In the present work, however, the boundary bias is fixed at its maximal value while the reservoir coupling $\Gamma$ is varied. Rather than controlling the onset of NDC, varying $\Gamma$ reveals that the doublon domain acts as a collective many-body energy barrier with interaction energy \begin{equation} E_{\mathrm{dom}} = U\sum_{j}\langle \hat{n}_{j\uparrow}\hat{n}_{j\downarrow}\rangle \simeq \frac{NU}{2}. \end{equation} Throughout this regime the density profile remains nearly unchanged over approximately one order of magnitude in $\Gamma$, indicating a remarkably robust correlated blockade. Physically, the boundary drive is insufficient to efficiently dissolve
the correlated doublon domain, while already being strong enough to
push the chain into the subdiffusive regime in which the domain
acts as an effective barrier to transport.

\paragraph*{High-drive resonance.} Upon further increasing the boundary driving, the correlated doublon domain gradually melts, leading to a partial recovery of the current. Numerically, the current reaches its maximum when \begin{equation} \Gamma_{\mathrm{high}} \simeq \frac{NU}{2}, \end{equation} coinciding with the interaction energy stored in the domain. The high-drive resonance therefore reflects the competition between two opposing mechanisms. For $\Gamma<\Gamma_{\mathrm{high}}$, increasing the boundary drive progressively melts the doublon domain, thereby enhancing charge transport. For $\Gamma>\Gamma_{\mathrm{high}}$, however, quantum Zeno suppression becomes dominant and the current decreases again. Importantly, the density profiles show that domain melting begins already below $\Gamma_{\mathrm{high}}$, indicating that the resonance does not mark the onset of melting itself but rather the crossover where the increasing Zeno suppression overtakes the transport enhancement arising from further domain dissolution. Motivated by Ref.~\cite{Hofman_Pothof}, which showed that in a closed extended Fermi-Hubbard chain a single doublon undergoes a damped oscillatory decay with a characteristic timescale scaling as $1/U$, we propose that $\Gamma_{high}$ corresponds to an optimal matching between the boundary driving rate and the characteristic relaxation timescale of the doublon-rich domain. Unlike the closed system of Ref.~\cite{Hofman_Pothof}, where the doublon population only partially decays due to the absence of dissipation, the boundary reservoirs in our setup provide an irreversible channel through which the interaction energy stored in the domain can be released. This phenomenological picture naturally suggests that the relevant energy scale increases with the total interaction energy of the domain, consistent with the observed scaling $\Gamma_{high} \sim NU/2$.

Both resonance positions admit natural physical interpretations. The
high-$\Gamma$ resonance corresponds to matching the drive rate to the
inverse timescale of collective domain reorganization.The low-$\Gamma$ resonance marks the onset of the subdiffusive
regime, which shifts to smaller $\Gamma$ with increasing $N$ and
is set by the superexchange scale $J_{\mathrm{ex}}$. A microscopic derivation of these scales and
their numerical prefactors is left for future work.

\section{\label{sec:dephasing}DEPHASING-ENHANCED TRANSPORT}

In this section, we investigate the influence of dephasing, another important environmental mechanism, on steady-state transport. In disorder-free non-interacting systems, dephasing destroys quantum coherence and drives a crossover from ballistic to diffusive transport, thereby reducing the steady-state current relative to the ballistic limit \cite{dephasing_diffusive}. In strongly interacting systems, however, its role becomes qualitatively different. Although dephasing is conventionally regarded as a source of decoherence, it can counterintuitively enhance transport in the presence of strong correlations. Such environment-assisted transport has previously been reported for strongly interacting spinless fermions \cite{PhysRevB.87.235130}. Here, we employ a similar framework to investigate the effect of local dephasing in the boundary-driven Fermi-Hubbard model.

\begin{figure}[htbp]
\centering
\includegraphics[width=.5\textwidth]{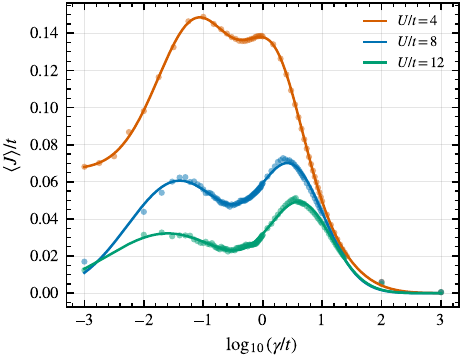}
\caption{\label{fig:jvsdeph}
Steady-state current as a function of the dephasing rate for an $N=8$ chain with $\Gamma=t$ for several interaction strengths.
}
\end{figure}

For non-interacting and weakly interacting systems, dephasing monotonically suppresses transport and no transport enhancement is observed. We therefore focus on the strongly interacting regime considered in Fig.~\ref{fig:jvsdeph}, where interactions qualitatively modify the transport behavior.

Figure~\ref{fig:jvsdeph} shows the steady-state current as a function of the dephasing rate $\gamma$ for several interaction strengths at fixed driving $\Gamma/t=1$. In striking contrast to the weakly interacting regime, strong interactions give rise to two distinct dephasing-induced transport enhancement peaks. Upon further increasing the dephasing strength, transport is eventually suppressed as the system enters the quantum Zeno regime, where the current asymptotically decays as $\langle J\rangle \propto 1/\gamma$.

To understand the microscopic origin of the two enhancement regimes, we examine the structure of the nonequilibrium steady state. The Hamiltonian conserves the numbers of spin-up and spin-down particles, whereas the boundary driving couples neighboring spin sectors according to
$(n_\uparrow,n_\downarrow)\rightarrow(n_\uparrow\pm1,n_\downarrow)$
and
$(n_\uparrow,n_\downarrow\pm1)$.
The many-body basis states are written as

\[
\ket{n_{1\downarrow},n_{2\downarrow},\ldots,n_{N\downarrow}}
\otimes
\ket{n_{1\uparrow},n_{2\uparrow},\ldots,n_{N\uparrow}},
\]
where $n_{i\sigma}\in\{0,1\}$ denotes the occupation of spin-$\sigma$ fermions on site $i$. States of the form
\[
\ket{1,x,0}\otimes\ket{1,y,0},
\]
where $x$ and $y$ denote arbitrary binary strings, are completely decoupled from the driving and therefore do not directly contribute to transport. These states are referred to as dark states. Their contribution is quantified by the dark-state weight operator

\[
\hat D
=
\hat n_{1\downarrow}
(1-\hat n_{N\downarrow})\hat n_{1\uparrow}
(1-\hat n_{N\uparrow}).
\]
Under strong boundary driving, the nonequilibrium steady state is dominated by doublon-rich configurations accumulated near the source reservoir (Sec.~\ref{sec:driving}), corresponding predominantly to high-energy eigenstates with large dark-state weight. Because transport out of these configurations requires either a particle to propagate across the entire chain to the drain or a hole to propagate in the opposite direction, strongly favoring the half-filled spin sector $(N/2,N/2)$ in the steady state. In the following, we therefore focus on this half-filled sector.

\begin{figure}[htbp]
\centering
\includegraphics[width=\textwidth]{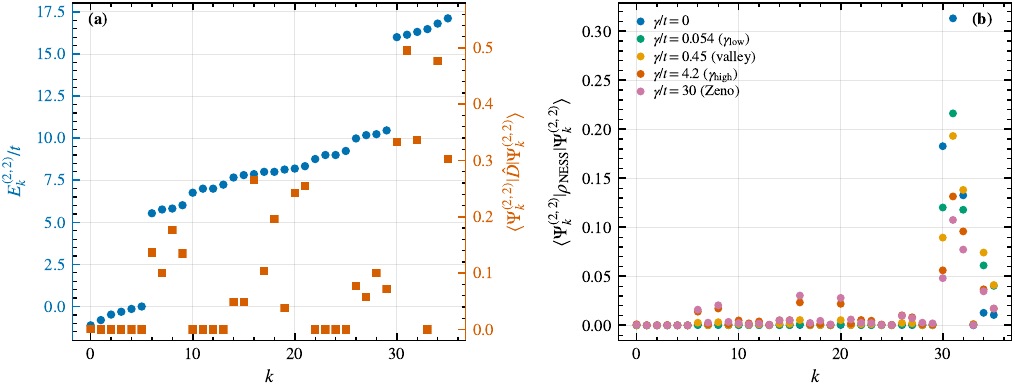}
\caption{\label{fig:fiveimp}(a)~Energy spectrum and dark-state weight $\langle\Psi_k^{(2,2)}|\hat{D}|\Psi_k^{(2,2)}\rangle$ of each eigenstate in the half-filled $(2,2)$ spin sector for an $N=4$ chain with $U/t=8$. (b)~Steady-state eigenstate populations $\langle\Psi_k^{(2,2)}|\rho_{\mathrm{NESS}}|\Psi_k^{(2,2)}\rangle$ for the representative dephasing rates indicated in the legend.
}
\end{figure}

Figure~\ref{fig:fiveimp}(a) displays the energy spectrum of the dominant half-filled sector and the corresponding dark-state weight of each eigenstate. In the spinless model of Ref.~\cite{PhysRevB.87.235130}, the highest-energy part of the spectrum is dominated by dark eigenstates associated with domain-pinned configurations, resulting in a single dephasing-induced enhancement peak. In the spinful Fermi-Hubbard model, by contrast, the combination of onsite interactions and spin degrees of freedom gives rise to a rich set of high-energy eigenstates within the uppermost band with substantially different dark-state weights, enabling two distinct enhancement regimes.

Since the local dephasing operators do not commute with the Hamiltonian, they induce incoherent transitions between Hamiltonian eigenstates. At weak dephasing, these transitions are predominantly confined within the uppermost band. As shown in Fig.~\ref{fig:fiveimp}(b), the steady-state population becomes redistributed among upper-band eigenstates toward those with smaller dark-state weight and higher mobility, giving rise to the first transport enhancement.

As dephasing increases further, transitions between different Hubbard bands become appreciable. Population is transferred from the highly occupied uppermost band into lower-energy bands containing eigenstates with significantly smaller dark-state weights, resulting in the second transport enhancement. Notably, comparing the optimal-dephasing and Zeno
regimes reveals an opposite state-selectivity: relative to the Zeno case,
the population at $\gamma_{\mathrm{high}}$ is enhanced in eigenstates with
vanishing dark-state weight in the intermediate bands [e.g., $k=10,12,22,23$
in Fig.~\ref{fig:fiveimp}], whereas in the Zeno regime the population is instead enhanced
in eigenstates with large dark-state weight [e.g., $k=8,16,20$]. Moderate dephasing thus preferentially channels population toward mobile
states, while strong dephasing locks it into dark eigenstates. Finally, for sufficiently strong dephasing, the suppression of coherent hopping through the quantum Zeno effect drives the system into the asymptotic regime where
\[
\langle J\rangle\propto\frac{1}{\gamma}.
\]

\begin{figure}[htbp]
\centering
\includegraphics[width=\textwidth]{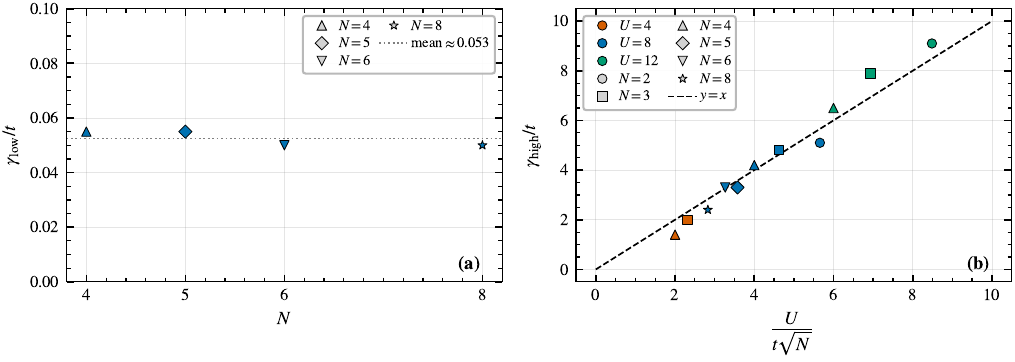}
\caption{\label{fig:dephasing_scaling}
Scaling of the two dephasing-induced enhancement peaks with system size
$N$ and interaction strength $U$, at fixed $\Gamma/t=1$. (a)~$\gamma_{\mathrm{low}}$ versus $N$ for $U/t=8$; the dotted line indicates
the mean value. (b)~$\gamma_{\mathrm{high}}$ versus $U/\sqrt{N}$, for
$N=2$--$8$ and $U/t=4,8,12$. The dashed line indicates $y=x$.
}
\end{figure}
The distinct microscopic origins of the two enhancement regimes are further supported by their different scaling behaviors. Fig.~\ref{fig:dephasing_scaling} summarizes the dependence of the optimal dephasing rates on the interaction strength and system size. The first enhancement peak occurs at relatively weak dephasing and is approximately independent of system size. Its position shifts toward
smaller dephasing rates with increasing $U$
(Fig.~\ref{fig:jvsdeph}), but no simple scaling law could
be identified within the parameter range considered. This behavior is consistent with the intra-band redistribution picture. 
At $\Gamma/t=1$, the boundary driving already induces transitions between 
many-body eigenstates even in the absence of dephasing, making it difficult 
to identify a single characteristic energy scale controlling the first peak. 
Nevertheless, the systematic shift of the peak position toward smaller 
dephasing rates with increasing $U$ is qualitatively consistent with the 
emergence of smaller intra-band energy splittings in the strongly interacting 
regime. In particular, virtual hopping processes can generate energy scales 
of order $t^2/U$ between strongly correlated states, which may influence the 
dephasing rate required to efficiently redistribute population within the 
upper Hubbard band.

In contrast, the second enhancement peak exhibits a robust scaling
behavior. As shown in Fig.~\ref{fig:dephasing_scaling}, the
corresponding optimal dephasing rate follows
\begin{equation}
\gamma_{\mathrm{high}} \sim \frac{U}{\sqrt{N}}.
\end{equation}
The linear dependence on $U$ is consistent with the inter-band origin
of the second enhancement, where the relevant energy scale is set by
the separation between Hubbard bands. The system-size dependence can
be understood as a competition between two transport processes:
dephasing-assisted depinning of the doublon-rich domain, which involves
an off-resonant transition with an energy cost of order $U$, and the
subsequent propagation of the resulting mobile configuration, which
becomes increasingly resistive with system size under strong dephasing.
As shown in Appendix~\ref{app:scaling_model}, a minimal two-process model
combining these contributions yields
$\gamma_{\mathrm{high}}\sim U/\sqrt{N}$.

\section{\label{sec:conclusion}CONCLUSIONS}

We have investigated steady-state transport in a boundary-driven one-dimensional
Fermi-Hubbard chain subject to local dephasing using numerical
solutions of the Lindblad master equation. Our results demonstrate that strong
on-site interactions qualitatively modify the role of both boundary driving and
dephasing in determining the steady-state current.

In the absence of dephasing, we find that the current exhibits a non-monotonic
dependence on the boundary driving rate $\Gamma$, with two well-separated
maxima whose positions scale differently with system size:
$\Gamma_{\mathrm{high}}\sim NU/2$, determined by the interaction energy scale
of doublon-rich configurations, and
$\Gamma_{\mathrm{low}}\sim 2J_{\mathrm{ex}}/N$, governed by the superexchange
energy scale. This separation reflects the coexistence of distinct charge and spin
energy scales in the strongly correlated regime.

Extending previous studies of dephasing-enhanced transport in spinless fermion
chains \cite{PhysRevB.87.235130}, we showed that local dephasing produces two
distinct transport-enhancement regimes in the spinful Fermi-Hubbard model. These
regimes arise from the interplay between on-site interactions and the spinful
many-body structure of the system. By analyzing the nonequilibrium steady state
in the Hamiltonian eigenbasis, we identified the first enhancement as
dephasing-induced redistribution within the upper Hubbard band toward eigenstates
with lower dark-state weight and higher mobility. The second enhancement results
from dephasing-induced transitions between Hubbard bands, with an optimal
dephasing rate following the scaling
$\gamma_{\mathrm{high}}\propto U/\sqrt{N}$, consistent with its inter-band
origin. At sufficiently strong dephasing, these enhancement mechanisms are
suppressed by the quantum Zeno effect, leading to the asymptotic decay
$\langle J\rangle\propto1/\gamma$.

Taken together, our results demonstrate that many-body correlations fundamentally
reshape the interplay between coherent transport, boundary driving, and
decoherence, producing transport phenomena absent in single-particle and
spinless models. While our results establish robust scaling behaviors for the characteristic driving and dephasing rates, a microscopic derivation of these scalings and their numerical prefactors from the underlying many-body dynamics remains an important direction for future work. The phenomenological picture developed here provides a physical interpretation of these trends, but a systematic treatment of the full many-body transition processes is needed to establish their microscopic origin. Extending this analysis to
higher-dimensional lattices may reveal additional transport regimes,
although the mechanism identified here relies in part on the distinct
charge ($U$) and spin (superexchange, $J_{\mathrm{ex}}$) energy scales that
emerge near half filling. Whether an analogous competition persists, and
how it manifests, in higher dimensions where spin and charge degrees of
freedom are not separated in the same manner as in one dimension remains
an open question. These predictions may be experimentally accessible with
current ultracold-atom platforms realizing boundary-driven Hubbard chains.

\appendix
\section{\label{app:scaling_model}Two-process model for the scaling of the second enhancement peak}

We develop a minimal model for the second dephasing-induced enhancement
peak. The physical picture is that transport out of the strongly
interacting, doublon-rich sector proceeds through two successive
processes. First, dephasing assists the depinning of a doublon-rich
domain by enabling an otherwise off-resonant hopping process with an
energy cost of order $U$. Once the domain is depinned, the resulting
mobile configuration propagates through the chain; this stage carries
no additional interaction-energy cost within the present picture and is
instead limited by dephasing-induced quantum-Zeno suppression. In the
steady state, the two processes carry the same particle flux in series,
so that their inverse rates can be viewed as current-limiting
resistances. The competition between these two contributions provides a
minimal mechanism for the observed system-size dependence of the optimal
dephasing rate.

\subsection{Dephasing-assisted depinning}

To illustrate the elementary depinning process, consider a
doublon-rich configuration in the uppermost Hubbard band. A
representative configuration can be written as
\begin{equation}
|D\rangle =
|\uparrow\downarrow,\uparrow\downarrow,0,0,\ldots\rangle .
\end{equation}
A nearest-neighbor hopping process across the domain boundary produces
a configuration such as
\begin{equation}
|D'\rangle =
|\uparrow\downarrow,\downarrow,\uparrow,0,\ldots\rangle .
\end{equation}
The two configurations differ by one doublon and therefore have an
energy difference of order $U$,
\begin{equation}
\Delta E
=
E_D-E_{D'}
\simeq U .
\end{equation}
The hopping Hamiltonian provides the coherent coupling between the two
configurations,
\begin{equation}
t_c
=
\left|\langle D'|\hat{H}_t|D\rangle\right|
\sim t .
\end{equation}
Thus, in the absence of dephasing, this process is off resonant in the
strongly interacting regime $t\ll U$.

The local dephasing operator is diagonal in the bare configuration
basis. For a site whose occupation changes during the hopping process,
\begin{equation}
\hat{n}_i|D\rangle
=
n_i|D\rangle,
\qquad
\hat{n}_i|D'\rangle
=
n_i'|D'\rangle ,
\end{equation}
with $n_i\neq n_i'$. Consequently, dephasing distinguishes the two
configurations and suppresses their coherence. The hopping matrix
element $t_c$ and the dephasing operator therefore play different roles:
the former provides the coherent transition amplitude, while the latter
broadens the off-resonant transition and enables an incoherent transfer
between the two configurations.

Within this two-level subspace, the dynamics is described by
\begin{equation}
\dot{\rho}
=
-i[\hat{H}_{\rm 2L},\rho]
+
\gamma\sum_i
\left(
\hat{n}_i\rho\hat{n}_i
-\frac{1}{2}
\left\{
\hat{n}_i^2,\rho
\right\}
\right),
\end{equation}
where
\begin{equation}
\hat{H}_{\rm 2L}
=
\begin{bmatrix}
E_D & -t_c\\
-t_c & E_{D'}
\end{bmatrix}.
\end{equation}
After subtracting an irrelevant constant from the Hamiltonian, this
can be written as
\begin{equation}
\hat{H}_{\rm 2L}
=
\begin{bmatrix}
0 & -t_c\\
-t_c & -\Delta E
\end{bmatrix}.
\end{equation}

For the choice $\hat{L}_i=\sqrt{\gamma}\hat{n}_i$ used in the main
model, the off-diagonal coherence decays at a rate proportional to
$\gamma$. Denoting the corresponding coherence-decay rate by
$\gamma_c$, the equations for coherence and population imbalance
are
\begin{align}
\dot{\rho}_{DD'}
&=
\left(
i\Delta E-\gamma_c
\right)
\rho_{DD'}
+
it_c\delta p ,
\label{eq:coherence_depin}
\\
\dot{\delta p}
&=
-4t_c\,\mathrm{Im}\rho_{DD'},
\label{eq:population_depin}
\end{align}
where
\begin{equation}
\delta p
=
p_{DD}-p_{D'D'} .
\end{equation}
The precise relation between $\gamma_c$ and the microscopic dephasing
rate $\gamma$ affects only numerical prefactors and is not relevant for
the scaling considered below.

When the coherence relaxes faster than the populations, it can be
adiabatically eliminated by setting
\begin{equation}
\dot{\rho}_{DD'}\simeq0 .
\end{equation}
This gives
\begin{equation}
\rho_{DD'}^{\rm ss}
=
\frac{it_c\delta p}
{\gamma_c-i\Delta E},
\end{equation}
and hence
\begin{equation}
\mathrm{Im}\rho_{DD'}^{\rm ss}
=
\frac{t_c\gamma_c}
{\gamma_c^2+\Delta E^2}
\delta p .
\end{equation}
Substituting this result into Eq.~\eqref{eq:population_depin} gives
\begin{equation}
\dot{\delta p}
=
-\Gamma_{\rm dep}(\gamma)\delta p ,
\end{equation}
with the effective depinning rate
\begin{equation}
\label{eq:depin_rate}
\Gamma_{\rm dep}(\gamma)
=
\frac{
4t_c^2\gamma_c
}{
\gamma_c^2+\Delta E^2
}.
\end{equation}
Since $t_c\sim t$ and $\Delta E\simeq U$, the relevant scaling is
therefore
\begin{equation}
\label{eq:depin_rate_scaling}
\boxed{
\Gamma_{\rm dep}(\gamma)
\propto
\frac{t^2\gamma}
{\gamma^2+U^2}
}.
\end{equation}

This expression has the limiting forms
\begin{align}
\Gamma_{\rm dep}(\gamma)
&\propto
\frac{t^2}{U^2}\gamma ,
&& \gamma\ll U ,
\\
\Gamma_{\rm dep}(\gamma)
&\propto
\frac{t^2}{\gamma},
&& \gamma\gg U .
\end{align}
Thus, weak dephasing assists the off-resonant depinning process,
whereas sufficiently strong dephasing suppresses it through the
quantum-Zeno mechanism. The depinning process considered in isolation
therefore has a characteristic optimal dephasing scale of order $U$.

In the driven steady state, depinning is not a single transient event:
the source continuously replenishes the doublon-rich configuration near
the injection boundary, which is then repeatedly depinned. Equation
\eqref{eq:depin_rate_scaling} therefore sets the characteristic rate at
which this stage can sustain particle flux. We consequently associate
with it an effective current-limiting resistance
\begin{equation}
\label{eq:R_dep}
R_{\rm dep}(\gamma)
\equiv
\Gamma_{\rm dep}^{-1}(\gamma)
\propto
\frac{\gamma^2+U^2}{\gamma}.
\end{equation}
Here we assume that replenishment of the doublon-rich configuration by
the source is not itself rate limiting, consistent with the fixed
driving rate considered in Sec.~\ref{sec:dephasing}.

\subsection{Transport after depinning}

Once the doublon-rich configuration has been depinned, the resulting
mobile population can propagate through the chain without repeatedly
paying the interaction-energy cost $U$. Within the present minimal
picture, this propagation is controlled by the hopping scale $t$ and by
dephasing rather than by the interaction gap.

The system-size dependence of this transport contribution can be
motivated from the corresponding noninteracting tight-binding problem.
For a boundary-driven chain with local dephasing, the steady-state
current has the form
\begin{equation}
\label{eq:exact_transport}
\langle J\rangle
=
\frac{a}{b+\gamma N},
\end{equation}
where $a$ and $b$ depend on the hopping amplitude and boundary coupling
but not on $N$ or $\gamma$ separately
\cite{review_theoretical,dephasing_diffusive}. In the regime where the
$\gamma$-dependent bulk contribution dominates, the inverse current
therefore contains a contribution
\begin{equation}
\label{eq:R_tr}
R_{\rm tr}(\gamma)
\propto
N\gamma.
\end{equation}
This linear dependence on $N$ reflects the Ohmic scaling of the
dephasing-induced diffusive bulk: local dephasing suppresses coherent
hopping through the quantum-Zeno mechanism, while the corresponding
bulk resistance increases linearly with the length of the chain.
Although we have motivated Eq.~\eqref{eq:exact_transport} from the
noninteracting limit, the same functional form is expected to hold in
interacting systems whenever dephasing dominates over all other energy
scales~\cite{review_theoretical}. Importantly, this contribution carries an
explicit factor of $N$, whereas the local depinning resistance in
Eq.~\eqref{eq:R_dep} does not.

\subsection{Optimal dephasing rate}

The two processes described above act sequentially and therefore
contribute additively to the current-limiting resistance,
\begin{equation}
\label{eq:R_total}
R(\gamma)
=
R_{\rm dep}(\gamma)
+
R_{\rm tr}(\gamma).
\end{equation}
Using Eqs.~\eqref{eq:R_dep} and~\eqref{eq:R_tr}, we write
\begin{equation}
\label{eq:R_total_scaling}
R(\gamma)
=
A\left(\gamma+\frac{U^2}{\gamma}\right)
+
BN\gamma, ,
\end{equation}
where $A$ and $B$ are dimensionless factors of order unity. Their
precise values depend on the microscopic details of the effective
depinning process and on the transport convention, respectively, and
are not needed for the scaling argument.

The optimal dephasing rate follows from minimizing the total
resistance,
\begin{equation}
\frac{\partial R}{\partial\gamma}=0 .
\end{equation}
This gives
\begin{equation}
A
\left(
1-\frac{U^2}{\gamma_{\rm high}^2}
\right)
+
BN
=
0 ,
\end{equation}
and therefore
\begin{equation}
\label{eq:gamma_opt_full}
\gamma_{\rm high}
=
\frac{U}
{\sqrt{1+(B/A)N}} .
\end{equation}
For sufficiently large $N$, Eq.~\eqref{eq:gamma_opt_full} reduces to
\begin{equation}
\boxed{
\gamma_{\rm high}
\propto
\frac{U}{\sqrt{N}} .
}
\end{equation}

The scaling thus follows from the competition between two distinct
processes. At weak dephasing, dephasing assists the depinning of the
doublon-rich domain by broadening an off-resonant transition with energy
mismatch of order $U$. At larger dephasing rates, the depinning process
becomes Zeno suppressed, while the subsequent propagation of the mobile
configuration is increasingly hindered by dephasing. Since the bulk
transport resistance grows linearly with system size, its contribution
becomes progressively more important as $N$ increases. The balance between these two contributions consequently shifts the
optimal dephasing rate from the local scale $\gamma\sim U$ toward
$\gamma_{\mathrm{high}}\sim U/\sqrt{N}$.

This construction is intended as a phenomenological two-process
description of the observed scaling rather than as a microscopic
derivation of the full many-body dynamics. The depinning step is
represented by a minimal two-level reduction, while the
system-size-dependent transport contribution is motivated by the
noninteracting transport result. The model therefore demonstrates how
the observed $U/\sqrt{N}$ dependence can emerge from the competition of
the two processes, while a microscopic treatment of the complete
many-body transition network would be required to derive the scaling
and its prefactor directly from the underlying Hubbard model.

\begin{acknowledgments}
The authors thank Ahmet Levent Subaşı for helpful discussions on tensor network methods.
\end{acknowledgments}

\bibliography{apssamp}

\begin{thebibliography}{29}%
\makeatletter
\providecommand \@ifxundefined [1]{%
 \@ifx{#1\undefined}
}%
\providecommand \@ifnum [1]{%
 \ifnum #1\expandafter \@firstoftwo
 \else \expandafter \@secondoftwo
 \fi
}%
\providecommand \@ifx [1]{%
 \ifx #1\expandafter \@firstoftwo
 \else \expandafter \@secondoftwo
 \fi
}%
\providecommand \natexlab [1]{#1}%
\providecommand \enquote  [1]{``#1''}%
\providecommand \bibnamefont  [1]{#1}%
\providecommand \bibfnamefont [1]{#1}%
\providecommand \citenamefont [1]{#1}%
\providecommand \href@noop [0]{\@secondoftwo}%
\providecommand \href [0]{\begingroup \@sanitize@url \@href}%
\providecommand \@href[1]{\@@startlink{#1}\@@href}%
\providecommand \@@href[1]{\endgroup#1\@@endlink}%
\providecommand \@sanitize@url [0]{\catcode `\\12\catcode `\$12\catcode `\&12\catcode `\#12\catcode `\^12\catcode `\_12\catcode `\%12\relax}%
\providecommand \@@startlink[1]{}%
\providecommand \@@endlink[0]{}%
\providecommand \url  [0]{\begingroup\@sanitize@url \@url }%
\providecommand \@url [1]{\endgroup\@href {#1}{\urlprefix }}%
\providecommand \urlprefix  [0]{URL }%
\providecommand \Eprint [0]{\href }%
\providecommand \doibase [0]{https://doi.org/}%
\providecommand \selectlanguage [0]{\@gobble}%
\providecommand \bibinfo  [0]{\@secondoftwo}%
\providecommand \bibfield  [0]{\@secondoftwo}%
\providecommand \translation [1]{[#1]}%
\providecommand \BibitemOpen [0]{}%
\providecommand \bibitemStop [0]{}%
\providecommand \bibitemNoStop [0]{.\EOS\space}%
\providecommand \EOS [0]{\spacefactor3000\relax}%
\providecommand \BibitemShut  [1]{\csname bibitem#1\endcsname}%
\let\auto@bib@innerbib\@empty
\bibitem [{\citenamefont {Dwiputra}\ and\ \citenamefont {Zen}(2021)}]{enaqt_mobilityedge}%
  \BibitemOpen
  \bibfield  {author} {\bibinfo {author} {\bibfnamefont {D.}~\bibnamefont {Dwiputra}}\ and\ \bibinfo {author} {\bibfnamefont {F.~P.}\ \bibnamefont {Zen}},\ }\bibfield  {title} {\bibinfo {title} {Environment-assisted quantum transport and mobility edges},\ }\href {https://doi.org/10.1103/PhysRevA.104.022205} {\bibfield  {journal} {\bibinfo  {journal} {Phys. Rev. A}\ }\textbf {\bibinfo {volume} {104}},\ \bibinfo {pages} {022205} (\bibinfo {year} {2021})}\BibitemShut {NoStop}%
\bibitem [{\citenamefont {Engel}\ \emph {et~al.}(2007)\citenamefont {Engel}, \citenamefont {Calhoun}, \citenamefont {Read}, \citenamefont {Ahn}, \citenamefont {Mančal}, \citenamefont {Cheng}, \citenamefont {Blankenship},\ and\ \citenamefont {Fleming}}]{fleming_fotosentez}%
  \BibitemOpen
  \bibfield  {author} {\bibinfo {author} {\bibfnamefont {G.~S.}\ \bibnamefont {Engel}}, \bibinfo {author} {\bibfnamefont {T.~R.}\ \bibnamefont {Calhoun}}, \bibinfo {author} {\bibfnamefont {E.~L.}\ \bibnamefont {Read}}, \bibinfo {author} {\bibfnamefont {T.-K.}\ \bibnamefont {Ahn}}, \bibinfo {author} {\bibfnamefont {T.}~\bibnamefont {Mančal}}, \bibinfo {author} {\bibfnamefont {Y.-C.}\ \bibnamefont {Cheng}}, \bibinfo {author} {\bibfnamefont {R.~E.}\ \bibnamefont {Blankenship}},\ and\ \bibinfo {author} {\bibfnamefont {G.~R.}\ \bibnamefont {Fleming}},\ }\bibfield  {title} {\bibinfo {title} {Evidence for wavelike energy transfer through quantum coherence in photosynthetic systems},\ }\href {https://doi.org/10.1038/nature05678} {\bibfield  {journal} {\bibinfo  {journal} {Nature}\ }\textbf {\bibinfo {volume} {446}},\ \bibinfo {pages} {782–786} (\bibinfo {year} {2007})}\BibitemShut {NoStop}%
\bibitem [{\citenamefont {Kassal}\ and\ \citenamefont {Aspuru-Guzik}(2012)}]{enaqt_ordered}%
  \BibitemOpen
  \bibfield  {author} {\bibinfo {author} {\bibfnamefont {I.}~\bibnamefont {Kassal}}\ and\ \bibinfo {author} {\bibfnamefont {A.}~\bibnamefont {Aspuru-Guzik}},\ }\bibfield  {title} {\bibinfo {title} {Environment-assisted quantum transport in ordered systems},\ }\href {https://doi.org/10.1088/1367-2630/14/5/053041} {\bibfield  {journal} {\bibinfo  {journal} {New Journal of Physics}\ }\textbf {\bibinfo {volume} {14}},\ \bibinfo {pages} {053041} (\bibinfo {year} {2012})}\BibitemShut {NoStop}%
\bibitem [{\citenamefont {León-Montiel}\ \emph {et~al.}(2015)\citenamefont {León-Montiel}, \citenamefont {Quiroz-Juárez}, \citenamefont {Quintero-Torres}, \citenamefont {Domínguez-Juárez}, \citenamefont {Moya-Cessa}, \citenamefont {Torres},\ and\ \citenamefont {Aragón}}]{León-Montiel_Quiroz-Juárez_Quintero-Torres_Domínguez-Juárez_Moya-Cessa_Torres_Aragón_2015}%
  \BibitemOpen
  \bibfield  {author} {\bibinfo {author} {\bibfnamefont {R.~d.}\ \bibnamefont {León-Montiel}}, \bibinfo {author} {\bibfnamefont {M.~A.}\ \bibnamefont {Quiroz-Juárez}}, \bibinfo {author} {\bibfnamefont {R.}~\bibnamefont {Quintero-Torres}}, \bibinfo {author} {\bibfnamefont {J.~L.}\ \bibnamefont {Domínguez-Juárez}}, \bibinfo {author} {\bibfnamefont {H.~M.}\ \bibnamefont {Moya-Cessa}}, \bibinfo {author} {\bibfnamefont {J.~P.}\ \bibnamefont {Torres}},\ and\ \bibinfo {author} {\bibfnamefont {J.~L.}\ \bibnamefont {Aragón}},\ }\bibfield  {title} {\bibinfo {title} {Noise-assisted energy transport in electrical oscillator networks with off-diagonal dynamical disorder},\ }\bibfield  {journal} {\bibinfo  {journal} {Scientific Reports}\ }\textbf {\bibinfo {volume} {5}},\ \href {https://doi.org/10.1038/srep17339} {10.1038/srep17339} (\bibinfo {year} {2015})\BibitemShut {NoStop}%
\bibitem [{\citenamefont {Maier}\ \emph {et~al.}(2019)\citenamefont {Maier}, \citenamefont {Brydges}, \citenamefont {Jurcevic}, \citenamefont {Trautmann}, \citenamefont {Hempel}, \citenamefont {Lanyon}, \citenamefont {Hauke}, \citenamefont {Blatt},\ and\ \citenamefont {Roos}}]{enaqt_10qubitnetwork}%
  \BibitemOpen
  \bibfield  {author} {\bibinfo {author} {\bibfnamefont {C.}~\bibnamefont {Maier}}, \bibinfo {author} {\bibfnamefont {T.}~\bibnamefont {Brydges}}, \bibinfo {author} {\bibfnamefont {P.}~\bibnamefont {Jurcevic}}, \bibinfo {author} {\bibfnamefont {N.}~\bibnamefont {Trautmann}}, \bibinfo {author} {\bibfnamefont {C.}~\bibnamefont {Hempel}}, \bibinfo {author} {\bibfnamefont {B.~P.}\ \bibnamefont {Lanyon}}, \bibinfo {author} {\bibfnamefont {P.}~\bibnamefont {Hauke}}, \bibinfo {author} {\bibfnamefont {R.}~\bibnamefont {Blatt}},\ and\ \bibinfo {author} {\bibfnamefont {C.~F.}\ \bibnamefont {Roos}},\ }\bibfield  {title} {\bibinfo {title} {Environment-assisted quantum transport in a 10-qubit network},\ }\href {https://doi.org/10.1103/PhysRevLett.122.050501} {\bibfield  {journal} {\bibinfo  {journal} {Phys. Rev. Lett.}\ }\textbf {\bibinfo {volume} {122}},\ \bibinfo {pages} {050501} (\bibinfo {year} {2019})}\BibitemShut {NoStop}%
\bibitem [{\citenamefont {Mohseni}\ \emph {et~al.}(2008)\citenamefont {Mohseni}, \citenamefont {Rebentrost}, \citenamefont {Lloyd},\ and\ \citenamefont {Aspuru-Guzik}}]{Lloyd_fotosentez}%
  \BibitemOpen
  \bibfield  {author} {\bibinfo {author} {\bibfnamefont {M.}~\bibnamefont {Mohseni}}, \bibinfo {author} {\bibfnamefont {P.}~\bibnamefont {Rebentrost}}, \bibinfo {author} {\bibfnamefont {S.}~\bibnamefont {Lloyd}},\ and\ \bibinfo {author} {\bibfnamefont {A.}~\bibnamefont {Aspuru-Guzik}},\ }\bibfield  {title} {\bibinfo {title} {Environment-assisted quantum walks in photosynthetic energy transfer},\ }\bibfield  {journal} {\bibinfo  {journal} {The Journal of Chemical Physics}\ }\textbf {\bibinfo {volume} {129}},\ \href {https://doi.org/10.1063/1.3002335} {10.1063/1.3002335} (\bibinfo {year} {2008})\BibitemShut {NoStop}%
\bibitem [{\citenamefont {Viciani}\ \emph {et~al.}(2016)\citenamefont {Viciani}, \citenamefont {Gherardini}, \citenamefont {Lima}, \citenamefont {Bellini},\ and\ \citenamefont {Caruso}}]{Viciani_Gherardini_Lima_Bellini_Caruso_2016a}%
  \BibitemOpen
  \bibfield  {author} {\bibinfo {author} {\bibfnamefont {S.}~\bibnamefont {Viciani}}, \bibinfo {author} {\bibfnamefont {S.}~\bibnamefont {Gherardini}}, \bibinfo {author} {\bibfnamefont {M.}~\bibnamefont {Lima}}, \bibinfo {author} {\bibfnamefont {M.}~\bibnamefont {Bellini}},\ and\ \bibinfo {author} {\bibfnamefont {F.}~\bibnamefont {Caruso}},\ }\bibfield  {title} {\bibinfo {title} {Disorder and dephasing as control knobs for light transport in optical fiber cavity networks},\ }\bibfield  {journal} {\bibinfo  {journal} {Scientific Reports}\ }\textbf {\bibinfo {volume} {6}},\ \href {https://doi.org/10.1038/srep37791} {10.1038/srep37791} (\bibinfo {year} {2016})\BibitemShut {NoStop}%
\bibitem [{\citenamefont {Zerah-Harush}\ and\ \citenamefont {Dubi}(2018)}]{enaqt_mechanism}%
  \BibitemOpen
  \bibfield  {author} {\bibinfo {author} {\bibfnamefont {E.}~\bibnamefont {Zerah-Harush}}\ and\ \bibinfo {author} {\bibfnamefont {Y.}~\bibnamefont {Dubi}},\ }\bibfield  {title} {\bibinfo {title} {Universal origin for environment-assisted quantum transport in exciton transfer networks},\ }\href {https://doi.org/10.1021/acs.jpclett.7b03306} {\bibfield  {journal} {\bibinfo  {journal} {The Journal of Physical Chemistry Letters}\ }\textbf {\bibinfo {volume} {9}},\ \bibinfo {pages} {1689} (\bibinfo {year} {2018})},\ \bibinfo {note} {pMID: 29537848},\ \Eprint {https://arxiv.org/abs/https://doi.org/10.1021/acs.jpclett.7b03306} {https://doi.org/10.1021/acs.jpclett.7b03306} \BibitemShut {NoStop}%
\bibitem [{\citenamefont {Zerah-Harush}\ and\ \citenamefont {Dubi}(2020)}]{NESSbio}%
  \BibitemOpen
  \bibfield  {author} {\bibinfo {author} {\bibfnamefont {E.}~\bibnamefont {Zerah-Harush}}\ and\ \bibinfo {author} {\bibfnamefont {Y.}~\bibnamefont {Dubi}},\ }\bibfield  {title} {\bibinfo {title} {Effects of disorder and interactions in environment assisted quantum transport},\ }\bibfield  {journal} {\bibinfo  {journal} {Physical Review Research}\ }\textbf {\bibinfo {volume} {2}},\ \href {https://doi.org/10.1103/PhysRevResearch.2.023294} {10.1103/PhysRevResearch.2.023294} (\bibinfo {year} {2020})\BibitemShut {NoStop}%
\bibitem [{\citenamefont {Rebentrost}\ \emph {et~al.}(2009)\citenamefont {Rebentrost}, \citenamefont {Mohseni}, \citenamefont {Kassal}, \citenamefont {Lloyd},\ and\ \citenamefont {Aspuru-Guzik}}]{ENAQT}%
  \BibitemOpen
  \bibfield  {author} {\bibinfo {author} {\bibfnamefont {P.}~\bibnamefont {Rebentrost}}, \bibinfo {author} {\bibfnamefont {M.}~\bibnamefont {Mohseni}}, \bibinfo {author} {\bibfnamefont {I.}~\bibnamefont {Kassal}}, \bibinfo {author} {\bibfnamefont {S.}~\bibnamefont {Lloyd}},\ and\ \bibinfo {author} {\bibfnamefont {A.}~\bibnamefont {Aspuru-Guzik}},\ }\bibfield  {title} {\bibinfo {title} {Environment-assisted quantum transport},\ }\href {https://doi.org/10.1088/1367-2630/11/3/033003} {\bibfield  {journal} {\bibinfo  {journal} {New Journal of Physics}\ }\textbf {\bibinfo {volume} {11}},\ \bibinfo {pages} {033003} (\bibinfo {year} {2009})}\BibitemShut {NoStop}%
\bibitem [{\citenamefont {Landi}\ \emph {et~al.}(2022)\citenamefont {Landi}, \citenamefont {Poletti},\ and\ \citenamefont {Schaller}}]{review_theoretical}%
  \BibitemOpen
  \bibfield  {author} {\bibinfo {author} {\bibfnamefont {G.~T.}\ \bibnamefont {Landi}}, \bibinfo {author} {\bibfnamefont {D.}~\bibnamefont {Poletti}},\ and\ \bibinfo {author} {\bibfnamefont {G.}~\bibnamefont {Schaller}},\ }\bibfield  {title} {\bibinfo {title} {Nonequilibrium boundary-driven quantum systems: Models, methods, and properties},\ }\href {https://doi.org/10.1103/RevModPhys.94.045006} {\bibfield  {journal} {\bibinfo  {journal} {Rev. Mod. Phys.}\ }\textbf {\bibinfo {volume} {94}},\ \bibinfo {pages} {045006} (\bibinfo {year} {2022})}\BibitemShut {NoStop}%
\bibitem [{\citenamefont {Bertini}\ \emph {et~al.}(2021)\citenamefont {Bertini}, \citenamefont {Heidrich-Meisner}, \citenamefont {Karrasch}, \citenamefont {Prosen}, \citenamefont {Steinigeweg},\ and\ \citenamefont {\ifmmode \check{Z}\else \v{Z}\fi{}nidari\ifmmode~\check{c}\else \v{c}\fi{}}}]{review_experiment}%
  \BibitemOpen
  \bibfield  {author} {\bibinfo {author} {\bibfnamefont {B.}~\bibnamefont {Bertini}}, \bibinfo {author} {\bibfnamefont {F.}~\bibnamefont {Heidrich-Meisner}}, \bibinfo {author} {\bibfnamefont {C.}~\bibnamefont {Karrasch}}, \bibinfo {author} {\bibfnamefont {T.}~\bibnamefont {Prosen}}, \bibinfo {author} {\bibfnamefont {R.}~\bibnamefont {Steinigeweg}},\ and\ \bibinfo {author} {\bibfnamefont {M.}~\bibnamefont {\ifmmode \check{Z}\else \v{Z}\fi{}nidari\ifmmode~\check{c}\else \v{c}\fi{}}},\ }\bibfield  {title} {\bibinfo {title} {Finite-temperature transport in one-dimensional quantum lattice models},\ }\href {https://doi.org/10.1103/RevModPhys.93.025003} {\bibfield  {journal} {\bibinfo  {journal} {Rev. Mod. Phys.}\ }\textbf {\bibinfo {volume} {93}},\ \bibinfo {pages} {025003} (\bibinfo {year} {2021})}\BibitemShut {NoStop}%
\bibitem [{\citenamefont {Bloch}\ \emph {et~al.}(2008)\citenamefont {Bloch}, \citenamefont {Dalibard},\ and\ \citenamefont {Zwerger}}]{experiment_coldgas}%
  \BibitemOpen
  \bibfield  {author} {\bibinfo {author} {\bibfnamefont {I.}~\bibnamefont {Bloch}}, \bibinfo {author} {\bibfnamefont {J.}~\bibnamefont {Dalibard}},\ and\ \bibinfo {author} {\bibfnamefont {W.}~\bibnamefont {Zwerger}},\ }\bibfield  {title} {\bibinfo {title} {Many-body physics with ultracold gases},\ }\href {https://doi.org/10.1103/RevModPhys.80.885} {\bibfield  {journal} {\bibinfo  {journal} {Rev. Mod. Phys.}\ }\textbf {\bibinfo {volume} {80}},\ \bibinfo {pages} {885} (\bibinfo {year} {2008})}\BibitemShut {NoStop}%
\bibitem [{\citenamefont {D'Errico}\ \emph {et~al.}(2013)\citenamefont {D'Errico}, \citenamefont {Moratti}, \citenamefont {Lucioni}, \citenamefont {Tanzi}, \citenamefont {Deissler}, \citenamefont {Inguscio}, \citenamefont {Modugno}, \citenamefont {Plenio},\ and\ \citenamefont {Caruso}}]{expansionofcoldatom}%
  \BibitemOpen
  \bibfield  {author} {\bibinfo {author} {\bibfnamefont {C.}~\bibnamefont {D'Errico}}, \bibinfo {author} {\bibfnamefont {M.}~\bibnamefont {Moratti}}, \bibinfo {author} {\bibfnamefont {E.}~\bibnamefont {Lucioni}}, \bibinfo {author} {\bibfnamefont {L.}~\bibnamefont {Tanzi}}, \bibinfo {author} {\bibfnamefont {B.}~\bibnamefont {Deissler}}, \bibinfo {author} {\bibfnamefont {M.}~\bibnamefont {Inguscio}}, \bibinfo {author} {\bibfnamefont {G.}~\bibnamefont {Modugno}}, \bibinfo {author} {\bibfnamefont {M.~B.}\ \bibnamefont {Plenio}},\ and\ \bibinfo {author} {\bibfnamefont {F.}~\bibnamefont {Caruso}},\ }\bibfield  {title} {\bibinfo {title} {Quantum diffusion with disorder, noise and interaction},\ }\href {https://doi.org/10.1088/1367-2630/15/4/045007} {\bibfield  {journal} {\bibinfo  {journal} {New Journal of Physics}\ }\textbf {\bibinfo {volume} {15}},\ \bibinfo {pages} {045007} (\bibinfo {year} {2013})}\BibitemShut {NoStop}%
\bibitem [{\citenamefont {Hess}(2019)}]{experiment_transport}%
  \BibitemOpen
  \bibfield  {author} {\bibinfo {author} {\bibfnamefont {C.}~\bibnamefont {Hess}},\ }\bibfield  {title} {\bibinfo {title} {Heat transport of cuprate-based low-dimensional quantum magnets with strong exchange coupling},\ }\href {https://doi.org/https://doi.org/10.1016/j.physrep.2019.02.004} {\bibfield  {journal} {\bibinfo  {journal} {Physics Reports}\ }\textbf {\bibinfo {volume} {811}},\ \bibinfo {pages} {1} (\bibinfo {year} {2019})},\ \bibinfo {note} {heat transport of cuprate-based low-dimensional quantum magnets with strong exchange coupling}\BibitemShut {NoStop}%
\bibitem [{\citenamefont {Schneider}\ \emph {et~al.}(2012)\citenamefont {Schneider}, \citenamefont {Hackermüller}, \citenamefont {Ronzheimer}, \citenamefont {Will}, \citenamefont {Braun}, \citenamefont {Best}, \citenamefont {Bloch}, \citenamefont {Demler}, \citenamefont {Mandt}, \citenamefont {Rasch},\ and\ \citenamefont {et~al.}}]{ultracold1}%
  \BibitemOpen
  \bibfield  {author} {\bibinfo {author} {\bibfnamefont {U.}~\bibnamefont {Schneider}}, \bibinfo {author} {\bibfnamefont {L.}~\bibnamefont {Hackermüller}}, \bibinfo {author} {\bibfnamefont {J.~P.}\ \bibnamefont {Ronzheimer}}, \bibinfo {author} {\bibfnamefont {S.}~\bibnamefont {Will}}, \bibinfo {author} {\bibfnamefont {S.}~\bibnamefont {Braun}}, \bibinfo {author} {\bibfnamefont {T.}~\bibnamefont {Best}}, \bibinfo {author} {\bibfnamefont {I.}~\bibnamefont {Bloch}}, \bibinfo {author} {\bibfnamefont {E.}~\bibnamefont {Demler}}, \bibinfo {author} {\bibfnamefont {S.}~\bibnamefont {Mandt}}, \bibinfo {author} {\bibfnamefont {D.}~\bibnamefont {Rasch}},\ and\ \bibinfo {author} {\bibnamefont {et~al.}},\ }\bibfield  {title} {\bibinfo {title} {Fermionic transport and out-of-equilibrium dynamics in a homogeneous hubbard model with ultracold atoms},\ }\href {https://doi.org/10.1038/nphys2205} {\bibfield  {journal} {\bibinfo  {journal} {Nature Physics}\ }\textbf {\bibinfo {volume} {8}},\ \bibinfo {pages} {213–218}
  (\bibinfo {year} {2012})}\BibitemShut {NoStop}%
\bibitem [{\citenamefont {Sologubenko}\ \emph {et~al.}(2000)\citenamefont {Sologubenko}, \citenamefont {Giann\'o}, \citenamefont {Ott}, \citenamefont {Ammerahl},\ and\ \citenamefont {Revcolevschi}}]{experiment_2}%
  \BibitemOpen
  \bibfield  {author} {\bibinfo {author} {\bibfnamefont {A.~V.}\ \bibnamefont {Sologubenko}}, \bibinfo {author} {\bibfnamefont {K.}~\bibnamefont {Giann\'o}}, \bibinfo {author} {\bibfnamefont {H.~R.}\ \bibnamefont {Ott}}, \bibinfo {author} {\bibfnamefont {U.}~\bibnamefont {Ammerahl}},\ and\ \bibinfo {author} {\bibfnamefont {A.}~\bibnamefont {Revcolevschi}},\ }\bibfield  {title} {\bibinfo {title} {Thermal conductivity of the hole-doped spin ladder system ${\mathrm{sr}}_{14\ensuremath{-}\mathit{x}}{\mathrm{ca}}_{\mathit{x}}{\mathrm{cu}}_{24}{O}_{41}$},\ }\href {https://doi.org/10.1103/PhysRevLett.84.2714} {\bibfield  {journal} {\bibinfo  {journal} {Phys. Rev. Lett.}\ }\textbf {\bibinfo {volume} {84}},\ \bibinfo {pages} {2714} (\bibinfo {year} {2000})}\BibitemShut {NoStop}%
\bibitem [{\citenamefont {Sologubenko}\ \emph {et~al.}(2007)\citenamefont {Sologubenko}, \citenamefont {Lorenz}, \citenamefont {Ott},\ and\ \citenamefont {Freimuth}}]{exp1}%
  \BibitemOpen
  \bibfield  {author} {\bibinfo {author} {\bibfnamefont {A.~V.}\ \bibnamefont {Sologubenko}}, \bibinfo {author} {\bibfnamefont {T.}~\bibnamefont {Lorenz}}, \bibinfo {author} {\bibfnamefont {H.~R.}\ \bibnamefont {Ott}},\ and\ \bibinfo {author} {\bibfnamefont {A.}~\bibnamefont {Freimuth}},\ }\bibfield  {title} {\bibinfo {title} {Thermal conductivity via magnetic excitations in spin-chain materials},\ }\href {https://doi.org/10.1007/s10909-007-9317-x} {\bibfield  {journal} {\bibinfo  {journal} {Journal of Low Temperature Physics}\ }\textbf {\bibinfo {volume} {147}},\ \bibinfo {pages} {387–403} (\bibinfo {year} {2007})}\BibitemShut {NoStop}%
\bibitem [{\citenamefont {Benenti}\ \emph {et~al.}(2009)\citenamefont {Benenti}, \citenamefont {Casati}, \citenamefont {Prosen}, \citenamefont {Rossini},\ and\ \citenamefont {\ifmmode \check{Z}\else \v{Z}\fi{}nidari\ifmmode~\check{c}\else \v{c}\fi{}}}]{PhysRevB.80.035110}%
  \BibitemOpen
  \bibfield  {author} {\bibinfo {author} {\bibfnamefont {G.}~\bibnamefont {Benenti}}, \bibinfo {author} {\bibfnamefont {G.}~\bibnamefont {Casati}}, \bibinfo {author} {\bibfnamefont {T.~c.~v.}\ \bibnamefont {Prosen}}, \bibinfo {author} {\bibfnamefont {D.}~\bibnamefont {Rossini}},\ and\ \bibinfo {author} {\bibfnamefont {M.}~\bibnamefont {\ifmmode \check{Z}\else \v{Z}\fi{}nidari\ifmmode~\check{c}\else \v{c}\fi{}}},\ }\bibfield  {title} {\bibinfo {title} {Charge and spin transport in strongly correlated one-dimensional quantum systems driven far from equilibrium},\ }\href {https://doi.org/10.1103/PhysRevB.80.035110} {\bibfield  {journal} {\bibinfo  {journal} {Phys. Rev. B}\ }\textbf {\bibinfo {volume} {80}},\ \bibinfo {pages} {035110} (\bibinfo {year} {2009})}\BibitemShut {NoStop}%
\bibitem [{\citenamefont {Mendoza-Arenas}\ \emph {et~al.}(2013)\citenamefont {Mendoza-Arenas}, \citenamefont {Grujic}, \citenamefont {Jaksch},\ and\ \citenamefont {Clark}}]{PhysRevB.87.235130}%
  \BibitemOpen
  \bibfield  {author} {\bibinfo {author} {\bibfnamefont {J.~J.}\ \bibnamefont {Mendoza-Arenas}}, \bibinfo {author} {\bibfnamefont {T.}~\bibnamefont {Grujic}}, \bibinfo {author} {\bibfnamefont {D.}~\bibnamefont {Jaksch}},\ and\ \bibinfo {author} {\bibfnamefont {S.~R.}\ \bibnamefont {Clark}},\ }\bibfield  {title} {\bibinfo {title} {Dephasing enhanced transport in nonequilibrium strongly correlated quantum systems},\ }\href {https://doi.org/10.1103/PhysRevB.87.235130} {\bibfield  {journal} {\bibinfo  {journal} {Phys. Rev. B}\ }\textbf {\bibinfo {volume} {87}},\ \bibinfo {pages} {235130} (\bibinfo {year} {2013})}\BibitemShut {NoStop}%
\bibitem [{\citenamefont {Lindblad}(1976)}]{Lindblad_1976b}%
  \BibitemOpen
  \bibfield  {author} {\bibinfo {author} {\bibfnamefont {G.}~\bibnamefont {Lindblad}},\ }\bibfield  {title} {\bibinfo {title} {On the generators of quantum dynamical semigroups},\ }\href {https://doi.org/10.1007/bf01608499} {\bibfield  {journal} {\bibinfo  {journal} {Communications in Mathematical Physics}\ }\textbf {\bibinfo {volume} {48}},\ \bibinfo {pages} {119–130} (\bibinfo {year} {1976})}\BibitemShut {NoStop}%
\bibitem [{\citenamefont {Breuer}\ and\ \citenamefont {Petruccione}(2010)}]{Breuer_Petruccione_2010}%
  \BibitemOpen
  \bibfield  {author} {\bibinfo {author} {\bibfnamefont {H.-P.}\ \bibnamefont {Breuer}}\ and\ \bibinfo {author} {\bibfnamefont {F.}~\bibnamefont {Petruccione}},\ }\href@noop {} {\emph {\bibinfo {title} {The theory of Open Quantum Systems}}}\ (\bibinfo  {publisher} {Clarendon},\ \bibinfo {year} {2010})\BibitemShut {NoStop}%
\bibitem [{\citenamefont {Santos}\ and\ \citenamefont {Landi}(2016)}]{micro_deriv}%
  \BibitemOpen
  \bibfield  {author} {\bibinfo {author} {\bibfnamefont {J.~P.}\ \bibnamefont {Santos}}\ and\ \bibinfo {author} {\bibfnamefont {G.~T.}\ \bibnamefont {Landi}},\ }\bibfield  {title} {\bibinfo {title} {Microscopic theory of a nonequilibrium open bosonic chain},\ }\bibfield  {journal} {\bibinfo  {journal} {Physical Review E}\ }\textbf {\bibinfo {volume} {94}},\ \href {https://doi.org/10.1103/physreve.94.062143} {10.1103/physreve.94.062143} (\bibinfo {year} {2016})\BibitemShut {NoStop}%
\bibitem [{\citenamefont {Zwolak}\ and\ \citenamefont {Vidal}(2004)}]{opentebd}%
  \BibitemOpen
  \bibfield  {author} {\bibinfo {author} {\bibfnamefont {M.}~\bibnamefont {Zwolak}}\ and\ \bibinfo {author} {\bibfnamefont {G.}~\bibnamefont {Vidal}},\ }\bibfield  {title} {\bibinfo {title} {Mixed-state dynamics in one-dimensional quantum lattice systems: A time-dependent superoperator renormalization algorithm},\ }\href {https://doi.org/10.1103/PhysRevLett.93.207205} {\bibfield  {journal} {\bibinfo  {journal} {Phys. Rev. Lett.}\ }\textbf {\bibinfo {volume} {93}},\ \bibinfo {pages} {207205} (\bibinfo {year} {2004})}\BibitemShut {NoStop}%
\bibitem [{\citenamefont {Dzhioev}\ and\ \citenamefont {Kosov}(2011)}]{superfermion}%
  \BibitemOpen
  \bibfield  {author} {\bibinfo {author} {\bibfnamefont {A.~A.}\ \bibnamefont {Dzhioev}}\ and\ \bibinfo {author} {\bibfnamefont {D.~S.}\ \bibnamefont {Kosov}},\ }\bibfield  {title} {\bibinfo {title} {Super-fermion representation of quantum kinetic equations for the electron transport problem},\ }\href {https://doi.org/10.1063/1.3548065} {\bibfield  {journal} {\bibinfo  {journal} {The Journal of Chemical Physics}\ }\textbf {\bibinfo {volume} {134}},\ \bibinfo {pages} {044121} (\bibinfo {year} {2011})},\ \Eprint {https://arxiv.org/abs/https://pubs.aip.org/aip/jcp/article-pdf/doi/10.1063/1.3548065/13788434/044121\_1\_online.pdf} {https://pubs.aip.org/aip/jcp/article-pdf/doi/10.1063/1.3548065/13788434/044121\_1\_online.pdf} \BibitemShut {NoStop}%
\bibitem [{\citenamefont {Fishman}\ \emph {et~al.}(2022)\citenamefont {Fishman}, \citenamefont {White},\ and\ \citenamefont {Stoudenmire}}]{itensor}%
  \BibitemOpen
  \bibfield  {author} {\bibinfo {author} {\bibfnamefont {M.}~\bibnamefont {Fishman}}, \bibinfo {author} {\bibfnamefont {S.~R.}\ \bibnamefont {White}},\ and\ \bibinfo {author} {\bibfnamefont {E.~M.}\ \bibnamefont {Stoudenmire}},\ }\bibfield  {title} {\bibinfo {title} {{The ITensor Software Library for Tensor Network Calculations}},\ }\href {https://doi.org/10.21468/SciPostPhysCodeb.4} {\bibfield  {journal} {\bibinfo  {journal} {SciPost Phys. Codebases}\ ,\ \bibinfo {pages} {4}} (\bibinfo {year} {2022})}\BibitemShut {NoStop}%
\bibitem [{\citenamefont {Prosen}\ and\ \citenamefont {\ifmmode \check{Z}\else \v{Z}\fi{}nidari\ifmmode~\check{c}\else \v{c}\fi{}}(2012)}]{Hightempdiffusive}%
  \BibitemOpen
  \bibfield  {author} {\bibinfo {author} {\bibfnamefont {T.~c.~v.}\ \bibnamefont {Prosen}}\ and\ \bibinfo {author} {\bibfnamefont {M.}~\bibnamefont {\ifmmode \check{Z}\else \v{Z}\fi{}nidari\ifmmode~\check{c}\else \v{c}\fi{}}},\ }\bibfield  {title} {\bibinfo {title} {Diffusive high-temperature transport in the one-dimensional hubbard model},\ }\href {https://doi.org/10.1103/PhysRevB.86.125118} {\bibfield  {journal} {\bibinfo  {journal} {Phys. Rev. B}\ }\textbf {\bibinfo {volume} {86}},\ \bibinfo {pages} {125118} (\bibinfo {year} {2012})}\BibitemShut {NoStop}%
\bibitem [{\citenamefont {Hofmann}\ and\ \citenamefont {Potthoff}(2012)}]{Hofman_Pothof}%
  \BibitemOpen
  \bibfield  {author} {\bibinfo {author} {\bibfnamefont {F.}~\bibnamefont {Hofmann}}\ and\ \bibinfo {author} {\bibfnamefont {M.}~\bibnamefont {Potthoff}},\ }\bibfield  {title} {\bibinfo {title} {Doublon dynamics in the extended fermi-hubbard model},\ }\href {https://doi.org/10.1103/PhysRevB.85.205127} {\bibfield  {journal} {\bibinfo  {journal} {Phys. Rev. B}\ }\textbf {\bibinfo {volume} {85}},\ \bibinfo {pages} {205127} (\bibinfo {year} {2012})}\BibitemShut {NoStop}%
\bibitem [{\citenamefont {Žnidarič}(2010)}]{dephasing_diffusive}%
  \BibitemOpen
  \bibfield  {author} {\bibinfo {author} {\bibfnamefont {M.}~\bibnamefont {Žnidarič}},\ }\bibfield  {title} {\bibinfo {title} {Dephasing-induced diffusive transport in the anisotropic heisenberg model},\ }\href {https://doi.org/10.1088/1367-2630/12/4/043001} {\bibfield  {journal} {\bibinfo  {journal} {New Journal of Physics}\ }\textbf {\bibinfo {volume} {12}},\ \bibinfo {pages} {043001} (\bibinfo {year} {2010})}\BibitemShut {NoStop}%
\end{thebibliography}%

\end{document}